\documentclass{PoS}

\newcommand{\be}{\begin{eqnarray}}
\newcommand{\ee}{\end{eqnarray}}

\def\e{\mathrm{e}}
\def\re{Re}
\def\conj#1{{{#1}^{*}}}
\def\mus{\hat\mu}
\def\xs{\hat{x}}
\def\ys{\hat{y}}

\def\K{\mathcal{K}}

\PoS{PoS(LAT2005)200}

\title{Dirac eigenvalue correlations in quenched QCD at finite density}
\ShortTitle{Dirac eigenvalue correlations in quenched QCD at finite density}

\author{\speaker{James C. Osborn}
        \thanks{Submitted to PoS on October 18, 2005.}\\
	Physics Department \& Center for Computational Science\\
        Boston University\\
	Boston, MA 02215, USA\\
        E-mail: \email{josborn@physics.bu.edu}}

\author{Tilo Wettig\\
        Institute for Theoretical Physics\\
        University of Regensburg\\
        93040 Regensburg, Germany\\
        E-mail: \email{tilo.wettig@physik.uni-regensburg.de}}

\abstract{
We compare eigenvalue correlations of the Dirac operator with a chemical
potential obtained from lattice simulations of quenched QCD with analytic
predictions obtained from chiral effective theories in the zero-momentum limit.
By comparing the density and two-point correlation function we show that
the analytic results agree with QCD at low energies.
We also examine the scale (Thouless energy) up to
which the zero-momentum approximation is valid.
}

\FullConference{XXIIIrd International Symposium on Lattice Field Theory\\
                25-30 July 2005\\
                Trinity College, Dublin, Ireland}

\begin{document}

\section{Introduction}

One of the hardest problems in lattice QCD has been the simulation of
QCD with a nonzero baryon chemical potential $\mu$.
Some progress has been made recently in lattice simulations with small chemical
potentials at temperatures around the chiral symmetry restoration transition
\cite{latmu},
however for large $\mu$ the main progress has come from analytic approaches.
Much of the analytic work at large $\mu$ has focused on the phase diagram
for $\mu$ larger than the critical $\mu_c$ for chiral symmetry restoration
\cite{phasemu}.
Here we will instead focus on the low-energy properties of QCD for
$\mu<\mu_c$ where chiral effective theories can provide exact results.

The low-energy eigenvalue density of quenched QCD with a chemical potential
was first derived from chiral effective partition functions using the
replica limit of the Toda lattice equation \cite{SV}.
The same density was later obtained by exact diagonalization of a chiral
random matrix theory (RMT) using the orthogonal polynomial method \cite{O}.
The agreement between the results that were derived from two different models
suggests that there is a common universality class to which both belong.
Comparisons to the quenched eigenvalue density for lattice QCD with a
chemical potential were carried out previously \cite{AW} using a slightly
different theoretical result \cite{A} that turned out to be not quite the
correct result for the chiral universality class.
A later comparison of the correct density to lattice
QCD data found very good agreement \cite{W}.

The solution of the RMT also provided results for all eigenvalue correlations
and was extended to the unquenched case.
The differences between the quenched and unquenched results have already
provided much insight into the nature of chiral symmetry breaking
in finite density QCD \cite{OSV}.
Here we look only at quenched eigenvalues since they are easily
obtained and save the comparison of unquenched eigenvalues for a
future work.
We note that recently comparisons of unquenched Dirac eigenvalues for
two-color QCD with lattice results were made \cite{nc2}.
For three-color QCD, all the low-energy eigenvalue correlations in a
finite volume $V$ with topological charge $\nu$ can be obtained from the
kernel
\be
\label{kernel}
\K(x,y) &=&
\frac{|\xs\ys|^{\nu+1}}{4 \pi \mus^2 (\xs\conj{\ys})^\nu}
\, \sqrt{
K_\nu\left(\frac{|\xs|^2}{4\mus^2}\right)
K_\nu\left(\frac{|\ys|^2}{4\mus^2}\right)
}
\,\e^{-\frac{\re(\xs^2+\ys^2)}{8\mus^2}}
\int_0^1 \e^{-2\mus^2 t} I_\nu(\xs\sqrt{t}) I_\nu(\conj{\ys}\sqrt{t})\,dt~~~~~
\ee
with $\xs=x\Sigma V$, $\ys=y\Sigma V$ and $\mus=\mu F \sqrt{V}$ \cite{O}.
The low-energy constants $\Sigma$ and $F$ are the tree level chiral condensate
and pion decay constant that appear in the chiral Lagrangian.
The density is then $\rho(z) = \K(z,z)$ and the two-point
correlation function used below is $Y_2(x,y)=|\K(x,y)|^2$.

Since (\ref{kernel}) is obtained from the zero-momentum limit of the
chiral Lagrangian, it is only valid
as long as the higher momentum modes can be neglected.
At $\mu=0$ this approximation is valid as long as the eigenvalues are
smaller than the scale $E_c \sim F^2/\Sigma\sqrt{V}$ \cite{OV}
which is the equivalent of the Thouless energy in condensed matter systems.
The average eigenvalue spacing at the origin for $\mu=0$ is known
to be $\Delta=\pi/\Sigma V$ \cite{levelsp} so that the dimensionless
Thouless energy (conductance) is $g \equiv E_c/\Delta \sim F^2\sqrt{V}$
(ignoring numerical factors).
Here we examine how well the analytic predictions for the eigenvalue
correlations agree with quenched lattice simulations,
identify the Thouless energy and see how it varies with $\mu$ and $V$.

\begin{table}[b]
\scriptsize
\begin{tabular}{|c|rrrrrrrr|rr|rr|}
\hline
size   &    $6^4$    & $6^4$  & $6^4$ & $6^4$ & $6^4$ & $6^4$ & $6^4$ & $6^4$ &
            $8^4$    & $8^4$  &
            $10^4$   & $10^4$ \\
\hline
$\mu a$ &   0.006    & 0.008  & 0.03  & 0.05  & 0.1   & 0.2   & 0.4 & 1.0 &
            0.003375 & 0.2    &
            0.00216  & 0.2    \\
\hline
configs  &  15500    & 508    & 19000 & 15802 & 15346 & 18500 & 15346 & 15342 &
            20000    & 18609  &
            4000     & 4000   \\
\hline
$a r_{max}$& 0.004    & 0.004  & 0.004 & 0.004 & 0.008 & 0.015 & - & - &
            0.001    & 0.005  &
            0.0004   & 0.001  \\
\hline
\end{tabular}
\normalsize
\caption{Lattice simulation and fitting parameters.}
\label{lattices}
\end{table}

\section{Determining the low-energy constants by fitting the density}

The lattices used for this study were all generated using the standard
plaquette action at lattice coupling $\beta=5.0$.
The eigenvalues were obtained from the staggered Dirac operator with different
chemical potentials.
The lattice sizes and values of $\mu$ used along with the number of
configurations studied are given in table \ref{lattices}.
The smallest value of $\mu$ for each volume was chosen to keep the value of
$\mu\sqrt{V}$ that appears in the analytic result (\ref{kernel}) constant.
More details on the choice of parameters and calculation of the eigenvalues
can be found in \cite{AW}.

\begin{figure}[t]
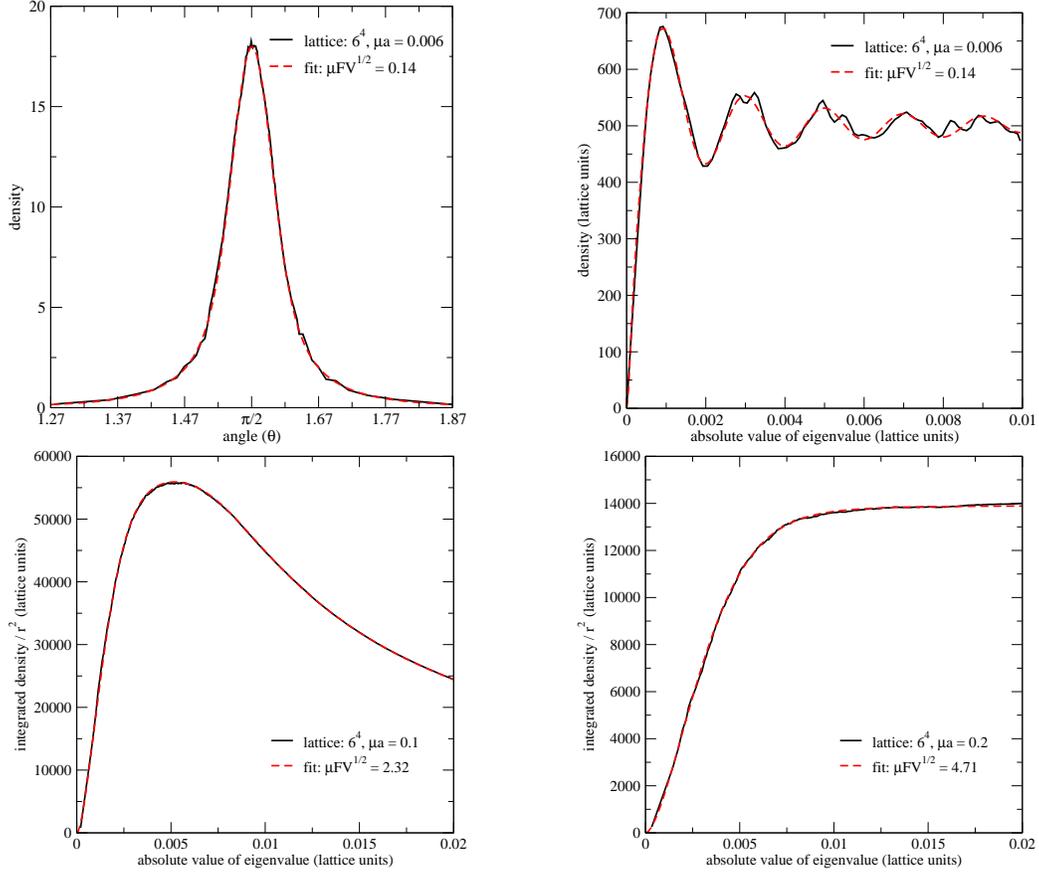

  \hfill
  \begin{minipage}[t]{.45\textwidth}
    \begin{center}
      \includegraphics[width=.9\columnwidth,clip]{dt5061.eps}
      \includegraphics[width=.9\columnwidth,clip]{dr5065.eps}
    \end{center}
  \end{minipage}
  \hfill
  \begin{minipage}[t]{.45\textwidth}
    \begin{center}
      \includegraphics[width=.9\columnwidth,clip]{dr5061.eps}
      \includegraphics[width=.9\columnwidth,clip]{dr5066.eps}
    \end{center}
  \end{minipage}
  \hfill
  \caption{Various visualizations of the eigenvalue density for quenched QCD
           on a $6^4$ lattice for different $\mu$.  The fits are from the
           analytic prediction (1.1).
           The fits for other lattice sizes and $\mu$ were also very good.}
  \label{d506x}
\end{figure}

First we need to determine the parameters $\Sigma$ and $F$ that appear in the
spectral correlations.
This is easily done by fitting to the eigenvalue density.
One approach is to fit the density along thin strips in the complex plane
that are essentially one dimensional as was done in \cite{AW}.
In order to use as much data as possible we take a different approach.
We fit to the integrated density
\be
I(r,\theta) = \int_{0}^{r} s\;ds \int_{0}^{\theta} d\phi
 ~ \rho(z=s\,\mathrm{e}^{i\phi}) ~.
\ee
To avoid complications involved with fitting to two variables we look at two
different one dimensional functions.
First is the radial dependence $I(r,\pi)$ which reduces to just the
integrated density along the imaginary axis when $\mu = 0$.
Here we only need to integrate to $\pi$ due to chiral symmetry.
We also look at the angular dependence given by $I(r_{max},\theta)$ with
$r_{max}$ fixed.  
We will compare the lattice data to the analytic expression at $\nu = 0$
since the staggered fermions at large lattice spacing do not see other
topological sectors.

In figure \ref{d506x} we show a few examples of the fits for various $\mu$.
In all cases the fits were obtained using the integrated density
just described, though for $\mu a=0.006$ we plot the density given by
either $d I(r_{max},\theta)/d\theta$ or $d I(r,\pi)/dr$ for better
illustration.  The values of $r_{max}$ we used in the angular dependence
(see table \ref{lattices}) are also the maximum values of $r$ we used
in the radial fits. 
This was determined by increasing $r_{max}$ until the $\chi^2/DOF$ of the
radial fit was around one.

At $\mu=0$ the eigenvalues are purely imaginary and the angular dependence
becomes a delta function at $\theta=\pi/2$.
The radial dependence clearly shows the characteristic oscillations of the
chiral RMT ensemble.
For small $\mu$ the angular dependence moves from a delta function to a
sharply peaked Gaussian while there is little change in the radial dependence.
As $\mu$ is increased more, the angular dependence becomes broader and
approaches a constant.
Meanwhile the radial dependence loses the oscillations as $\mu$ is increased.

\begin{figure}[t]
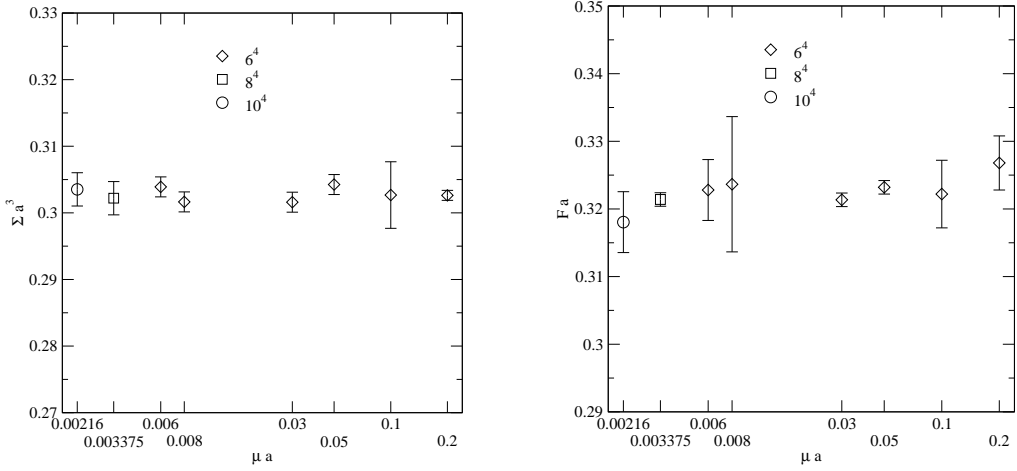

  \hfill
  \begin{minipage}[t]{.45\textwidth}
    \begin{center}
      \includegraphics[width=.9\columnwidth,clip]{sigmamu.eps}
    \end{center}
  \end{minipage}
  \hfill
  \begin{minipage}[t]{.45\textwidth}
    \begin{center}
      \includegraphics[width=.9\columnwidth,clip]{fmu.eps}
    \end{center}
  \end{minipage}
  \hfill
  \vspace{-3mm}
  \caption{Fitted values of the low-energy constants $\Sigma$ and $F$ for
           the different values of $\mu$ and $V$.
           All the fitted values are consistent with statistical errors
           between 0.3--3\%.}
  \vspace{-2mm}
  \label{lecmu}
\end{figure}

Since we are using staggered fermions we actually have four ``tastes''
in the simulation instead of just one flavor.
To correct for this we can simply replace $V$ in (\ref{kernel}) with $4V$
 \cite{DHSS}.
In that paper the authors also demonstrate how to obtain values for $\Sigma$ and $F$,
however they use the eigenvalue correlation function for imaginary isospin
chemical potential.
In figure \ref{lecmu} we show the obtained fit values for the different
volumes and values of $\mu$ used.
We cannot obtain independent fits for $\Sigma$ and $F$ for all values
of $\mu$ because in the large-$\hat\mu$ limit the density reduces to a
function only of the ratio $\Sigma/F$ \cite{A}.
All the values of $\Sigma$ and $F$ that we were able to obtain
are consistent with each other.

\section{Finding the Thouless energy by examining the eigenvalue correlations}

Now that the parameters have been fixed, we look at the eigenvalue correlations
to see what the effective range of the zero-momentum theory is.
One convenient quantity to use is the number variance.
For complex eigenvalues this can be defined for some region $A$ of the complex
plane as
\be
\Sigma_2(r) = r - \int_{A} d^2 z_1 \int_{A} d^2 z_2 ~ Y_2(z_1,z_2)
\ee
with $r=\int_{A} d^2 z \,\rho(z)$ \cite{GNV}.
Once the values of $\Sigma$ and $F$ are set there are no more free parameters.
Here we take $A$ to be circles centered on the origin with varying radius.

\begin{figure}[t]
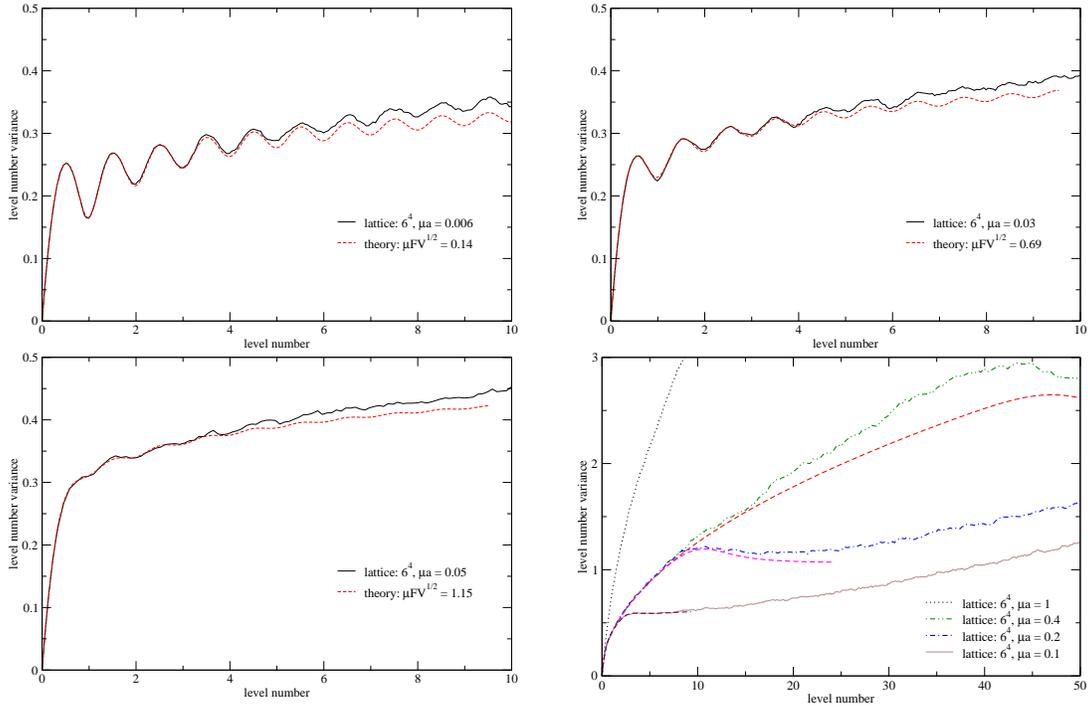

  \hfill
  \begin{minipage}[t]{.45\textwidth}
    \begin{center}
      \includegraphics[width=\columnwidth,clip]{nr5061.eps}
      \includegraphics[width=\columnwidth,clip]{nr5064.eps}
    \end{center}
  \end{minipage}
  \hfill
  \begin{minipage}[t]{.45\textwidth}
    \begin{center}
      \includegraphics[width=\columnwidth,clip]{nr5063.eps}
      \includegraphics[width=\columnwidth,clip]{nr5065-8.eps}
    \end{center}
  \end{minipage}
  \hfill
  \caption{Number variance $\Sigma_2(r)$ versus the average number of
           eigenvalues $r$ in a circle of varying size around the origin.
           The dashed lines are the analytic curves
           using the values of the low-energy constants obtained from the
           density fits.}
  \vspace{-2mm}
  \label{nr506x}
\end{figure}

In figure \ref{nr506x} we show the number variance versus level number ($r$
above) on $6^4$ lattices over a range of values of $\mu$.
For all values of $\mu a \le 0.4$ we see very good agreement between the
lattice results and the theoretical prediction up to the scale of a
few eigenvalues.
For smaller $\mu$ distinctive oscillations are clearly seen in both the lattice
data and theoretical result.
As $\mu$ increases the oscillations go away and the number variance approaches
a curve given by taking $\hat\mu$ to infinity in (\ref{kernel}).
This can be seen in the lower right plot of figure \ref{nr506x} for values
of $\mu a = 0.1$, $0.2$ and $0.4$ which all agree for small values of $r$ and
eventually turn away.
However for $\mu a = 1$ the number variance jumps up dramatically and does
not agree with the theory.
This is most likely due to saturation of the particle number on the finite
lattice.
Here we expect the eigenvalue statistics to move from RMT to Poisson
as was seen in \cite{MPW}.
By estimating the Thouless energy as the point where the number variance
on the lattice begins to deviate from the theoretical curve, we can
see that it increases with $\mu$ from about 3 to about 10 eigenvalues.
The precise form of the $\mu$-dependence needs  further study.

\begin{figure}[t]
  \hfill
  \begin{minipage}[t]{.45\textwidth}
    \begin{center}
      \includegraphics[width=\columnwidth,clip]{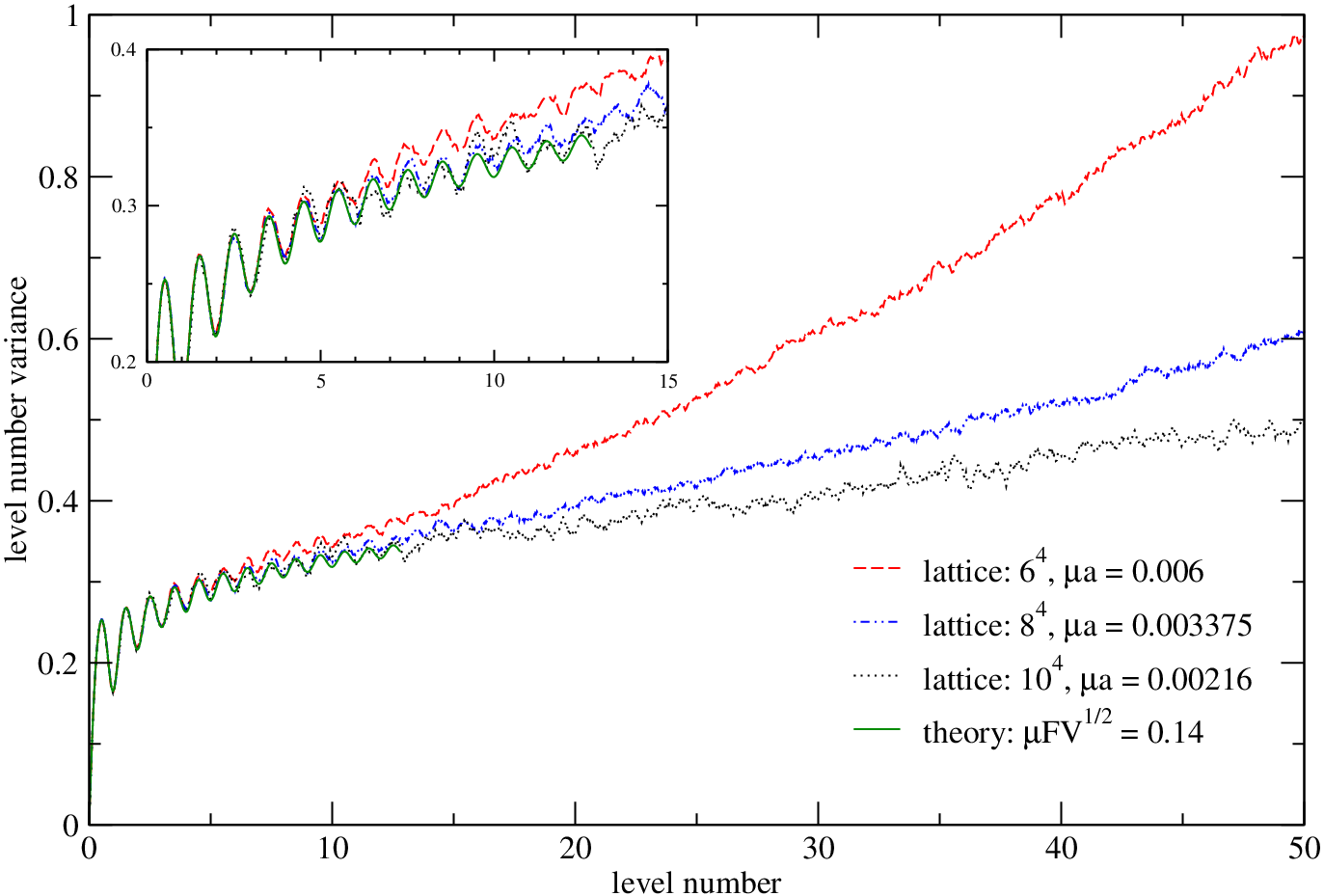}
      \caption{Number variance for small $\mu a$ and different $V$.}
      \label{nr50x1}
    \end{center}
  \end{minipage}
  \hfill
  \begin{minipage}[t]{.45\textwidth}
    \begin{center}
      \includegraphics[width=\columnwidth,clip]{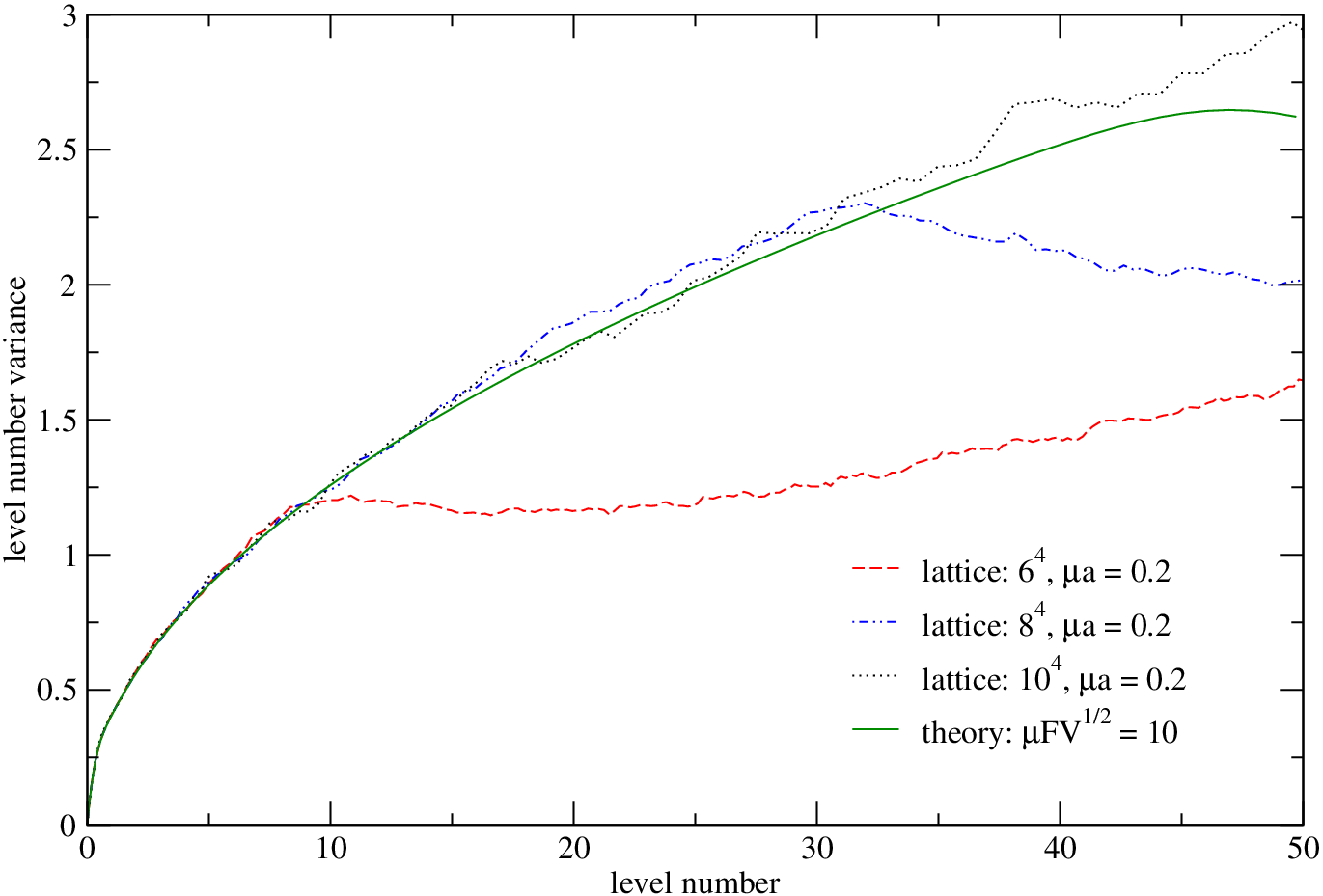}
      \caption{Number variance for large $\mu a = 0.2$ and different $V$.}
      \label{nr50x6}
    \end{center}
  \end{minipage}
  \hfill
\end{figure}

Next we look at the scaling of the Thouless energy with $V$.
In figure \ref{nr50x1} we show the number variance for the smallest values of
$\mu$ for each volume.
Here we clearly see that the Thouless energy is increasing with
the volume.
We do not have enough statistics at the largest volume to accurately measure
the exact scaling, but it does seem to be consistent with the $\sqrt{V}$
behavior expected at $\mu=0$.
For larger $\mu a=0.2$ (see figure \ref{nr50x6}), the Thouless energy also
appears to increase 
with volume but again we do not yet have enough statistics for an accurate
determination of the scaling.

\section{Conclusions}

We have shown that the low-energy eigenvalue density and two-point
correlations of quenched QCD with a chemical potential are accurately
determined from the chiral effective theory.
From fitting the density to lattice calculations we easily obtained
the low-energy constants $\Sigma$ and $F$. 
We also could clearly see the Thouless energy and its general dependence
on $V$ and $\mu$.  In future work, this dependence should be
quantified.  We should also move to weaker coupling and larger lattices.

We would like to thank J.J.M. Verbaarschot for helpful discussions that inspired
this work.

\end{document}